\documentclass[conference]{IEEEtran}
\IEEEoverridecommandlockouts

\usepackage{cite}
\usepackage{amsmath,amssymb,amsfonts}
\usepackage{algorithmic}
\usepackage{algorithm}
\usepackage{graphicx}
\usepackage{textcomp}
\usepackage{xcolor}
\usepackage{booktabs}
\usepackage{amsthm}

\usepackage{multirow}
\usepackage{url}
\usepackage{hyperref}
\usepackage{balance}
\newtheorem{theorem}{Theorem}

\def\BibTeX{{\rm B\kern-.05em{\sc i\kern-.025em b}\kern-.08em
    T\kern-.1667em\lower.7ex\hbox{E}\kern-.125emX}}

\begin{document}

\title{SecureDrive-FL: Joint Differential Privacy and\\
Gradient-Aware Selective Homomorphic Encryption\\
for Federated Driver Monitoring}

\makeatletter
\newcommand{\linebreakand}{%
  \end{@IEEEauthorhalign}
  \hfill\mbox{}\par
  \mbox{}\hfill\begin{@IEEEauthorhalign}
}

\author{%
\IEEEauthorblockN{Baran Can Gül\IEEEauthorrefmark{1},
Hanuma Siddhartha Tunuguntla\IEEEauthorrefmark{2},
Anjana Arvind Naik\IEEEauthorrefmark{1},
Abhishek Vijay Potekar\IEEEauthorrefmark{1},
\linebreakand
Nasser Jazdi\IEEEauthorrefmark{1},
and Michael Weyrich\IEEEauthorrefmark{1}}%
\IEEEauthorblockA{\IEEEauthorrefmark{1}Institute of Industrial Automation and Software Engineering, University of Stuttgart, Stuttgart, Germany}%
\IEEEauthorblockA{\IEEEauthorrefmark{2}EngD Automotive Systems Design, Eindhoven University of Technology (TU/e), Eindhoven, The Netherlands}%
\IEEEauthorblockA{Email: baran-can.guel@ias.uni-stuttgart.de,
h.s.tunuguntla@tue.nl,
st189825@stud.uni-stuttgart.de,
\linebreakand
st191608@stud.uni-stuttgart.de,
nasser.jazdi@ias.uni-stuttgart.de,
michael.weyrich@ias.uni-stuttgart.de}%
}

\maketitle

\begin{abstract}
Federated Learning (FL) enables privacy-aware distributed training, yet gradient updates remain exploitable: Man-in-the-Middle (MitM) interception exposes updates in transit, while model poisoning corrupts global convergence. We first introduce \textbf{GASHE} (Gradient-Aware Selective Homomorphic Encryption), a novel selective encryption strategy that dynamically identifies and encrypts only the gradient components exceeding a DP-calibrated sensitivity threshold, rather than encrypting all parameters uniformly as in static layer-based or full-parameter CKKS schemes. Building on GASHE, we introduce \textbf{SecureDrive-FL}, a federated driver monitoring framework that couples DP-SGD with GASHE to create the first closed-loop DP\,+\,HE privacy pipeline: DP-SGD calibration parameters directly derive the GASHE encryption mask, unifying training-time privacy and communication-time confidentiality. Evaluated on a ten-class distracted driver classification task under non-IID federated splits, SecureDrive-FL matches DP-SGD alone's poisoning resistance ($73.6\%$ vs.\ $74.0\%$ accuracy, $3.9\%$ Attack Success Rate for both) while additionally withstanding MitM interception, where DP-SGD alone collapses to near-random accuracy ($78.2\%$ vs.\ $10.4\%$), all under only ${\approx}8$--$10\%$ additional runtime overhead relative to DP-SGD alone---under DP-SGD noise injection with per-round privacy parameter $\varepsilon_0{=}4$.    
\end{abstract}


\section{Introduction}
\label{sec:intro}
The rapid growth of Internet of Things (IoT) devices and connected vehicles has led to the generation of large-scale distributed datasets covering driver behavior analysis, intelligent transportation systems, and autonomous driving~\cite{guel2023}. Federated Learning (FL) has emerged as a compelling paradigm to exploit these distributed datasets without centralizing sensitive raw data: local models are trained on edge devices and only gradient updates are transmitted to a central aggregator~\cite{mcmahan2017communication}. Yet, FL provides no formal privacy guarantee by itself. A growing body of evidence shows that adversaries can reconstruct sensitive information from gradient updates through model inversion~\cite{fredrikson2015model}, membership inference~\cite{shokri2017membership}, GAN-based reconstruction~\cite{hitaj2017deep}, and deep gradient leakage~\cite{zhu2019deep}. The threat surface is further expanded in automotive deployments, where model updates traverse shared cloud infrastructure operated by OEMs, insurers, and third-party service providers, each representing a potential interception point~\cite{gul2024fed}.

Two primary families of Privacy-Enhancing Technologies (PETs) have been studied independently for FL. Fully Homomorphic Encryption (FHE) allows the server to aggregate encrypted updates without ever decrypting them~\cite{gentry2009fhe}, but encrypting every model parameter incurs prohibitive memory and latency overhead on resource-constrained edge devices~\cite{gupta2022memfhe}. Differential Privacy (DP), realized via DP-SGD~\cite{abadi2016deep}, provides rigorous, quantifiable privacy guarantees by injecting calibrated Gaussian noise into gradient updates; however, the well-known privacy-utility trade-off means that high privacy levels substantially degrade model accuracy~\cite{geyer2017dpfl}.

Prior hybrid works~\cite{truex2020hybrid,zhang2020batchcrypt,hijazi2023secure} treat DP and HE as largely independent, orthogonally applied modules. In contrast, our key insight is to derive the encryption decision directly from the DP-SGD calibration itself: the clipping norm $C$ and noise scale $\sigma$ used in DP-SGD already identify which gradients dominate the DP noise floor and are therefore most informative, and most dangerous if transmitted in plaintext. This DP-informed gradient sensitivity threshold is used as the GASHE mask, creating a principled, closed-loop coupling between training-time privacy and communication-time confidentiality. The specific contributions of this work are:
\begin{itemize}
  \item \textbf{GASHE:} We introduce Gradient-Aware Selective Homomorphic Encryption, a novel, architecture-agnostic selective encryption strategy that dynamically targets CKKS encryption to the $k$ gradient components whose magnitudes exceed the DP-calibrated threshold $\tau_i$, retaining formal privacy semantics without encrypting the full parameter vector. GASHE can be deployed as a standalone communication-security primitive, independently of DP-SGD.

  \item \textbf{DP-GASHE coupling (SecureDrive-FL):} To the best of our knowledge, SecureDrive-FL is the first framework to explicitly derive HE encryption decisions from DP-SGD sensitivity bounds, creating a unified privacy pipeline with DP-SGD noise injection at training time (per-round $\varepsilon_0{=}4$, $\sigma{=}1.2112$, $C{=}1.2$, $\delta{=}10^{-5}$) and IND-CPA confidentiality at communication time (Theorem~\ref{thm:indcpa}).
  
  \item \textbf{Multi-threat robustness evaluation:} We evaluate SecureDrive-FL against model poisoning and MitM gradient interception, showing that the joint DP+GASHE design closes the MitM attack surface while matching DP-SGD alone's poisoning resistance. Under poisoning, SecureDrive-FL attains accuracy ($73.6\%$ vs.\ $74.0\%$), Macro-F1 ($73.3\%$ vs.\ $73.7\%$), and Attack Success Rate ($3.9\%$ vs.\ $3.9\%$) comparable to DP-SGD alone, with a marginal Macro-Precision gain ($75.9\%$ vs.\ $75.7\%$), while additionally preventing the accuracy collapse DP-SGD alone suffers under MitM ($78.2\%$ vs.\ $10.4\%$). Undefended FedAvg and FL+GASHE-only, by contrast, both collapse under poisoning (${\approx}10$--$14\%$ accuracy), so their nominally low ASR reflects training failure, not robustness.

  \item \textbf{Automotive use-case validation:} We demonstrate practical feasibility on a ten-class distracted driver recognition task with non-IID, user-partitioned FL and lightweight MobileNetV2-Tiny, achieving ${\approx}74\%$ accuracy under active attack with DP-SGD configured at per-round $\varepsilon_0{=}4$.
\end{itemize}

The remainder of the paper is organized as follows: Section~\ref{sec:related} reviews related work, Section~\ref{sec:framework} presents the SecureDrive-FL framework, and Section~\ref{sec:setup} details the threat model and our adversarial evaluation methodology. Section~\ref{sec:results} describes the implementation and experimental results and Section~\ref{sec:conclusion} provides the summary of this work.

\section{Related Work}
\label{sec:related}
We review privacy threats in FL, DP and HE mechanisms, and hybrid robustness-oriented frameworks, then position SecureDrive-FL against their limitations.
\subsection{Privacy Threats in Federated Learning}
Standard FL is vulnerable to a range of inference and interception attacks that exploit the shared model updates exchanged between clients and the server. Model inversion attacks demonstrated in~\cite{fredrikson2015model} exploit model confidence scores to reconstruct sensitive input attributes, while membership inference attacks introduced in~\cite{shokri2017membership} reveal whether specific records were part of the training set. GAN-based information leakage in collaborative learning is analyzed in~\cite{hitaj2017deep}, and unintended feature leakage through shared representations is examined in~\cite{melis2019exploiting}. Deep gradient leakage is investigated in~\cite{zhu2019deep}, showing that raw training samples can often be reconstructed directly from gradient updates under standard FL. More recent analyses of gradient leakage in federated environments are provided in~\cite{arXiv2510gradleak}. These works collectively establish that data locality alone is insufficient: exchanged gradients remain an exploitable attack surface that can be targeted both passively (eavesdropping) and actively (manipulation). 

\subsection{Differential Privacy in Federated Learning}
Differential Privacy provides a formal framework for bounding the information that individual records contribute to model outputs~\cite{dwork2006dp}. The DP-SGD mechanism proposed in~\cite{abadi2016deep} realizes this in deep learning via per-example gradient clipping and calibrated Gaussian noise injection, achieving formal $(\varepsilon,\delta)$-DP guarantees. Its adaptation to the federated setting is studied in~\cite{mcmahan2018dpfl}, where client-level DP is enforced across communication rounds, and a client-level perspective with adaptive clipping is developed in~\cite{geyer2017dpfl} to reduce utility degradation. Tighter privacy accounting using R\'enyi DP is introduced in~\cite{mironov2017renyi} and extended in~\cite{bun2016cdp}, enabling more efficient budget management over many rounds.

Despite these advances, a key limitation remains: as shown in~\cite{zhu2019deep}, gradient inversion can partially succeed even in the presence of DP noise when the signal-to-noise ratio is favorable, particularly for small batch sizes. Furthermore, DP-SGD operates only during local training and provides no confidentiality guarantee for the gradient updates transmitted over the network, leaving them exposed to Man-in-the-Middle interception. 

\subsection{Homomorphic Encryption in Federated Learning}
Fully Homomorphic Encryption, initially constructed in~\cite{gentry2009fhe}, allows computation on encrypted data and has been adopted for secure aggregation in FL. BatchCrypt~\cite{zhang2020batchcrypt} reduces HE overhead via quantization and batching; secret-sharing-based secure aggregation~\cite{bonawitz2017secagg}
enables servers to recover only the aggregate; and TenSEAL~\cite{benaissa2021tenseal} with~\cite{hijazi2023secure} bring CKKS-based encryption to general ML and IoT FL pipelines respectively.

However, all of the above approaches suffer from a critical practical limitation: they encrypt \emph{all} model parameters uniformly, incurring computational and memory overhead that is prohibitive on resource-constrained edge devices. Partial or layer-based encryption schemes attempt to reduce this cost by encrypting only selected layers, but they rely on static, architecture-dependent rules that do not adapt to the actual privacy sensitivity of individual gradient components during training. 

\subsection{Hybrid Privacy Frameworks and Robustness}
A comprehensive survey of FL attacks and defenses is provided in~\cite{lyu2022privacy}, and strong multi-round poisoning attacks circumventing many existing defenses are introduced in~\cite{xie2025poisonedfl}. Hybrid DP--HE approaches~\cite{truex2020hybrid} apply noise-based privacy and encrypted aggregation as independently configured modules,  without directing encryption toward the most sensitive gradients.  A representative hybrid defense~\cite{truex2019hybrid} combines DP and SMC to improve accuracy over single-mechanism baselines, but DP and SMC are configured independently: the DP calibration does not inform which updates SMC protects, and robustness to poisoning is not evaluated. SecureDrive-FL instead derives the HE mask directly from the DP-SGD clipping norm $C$ and noise scale $\sigma$, and evaluates the resulting pipeline against both inference and poisoning adversaries in a realistic automotive FL setting.

\section{SecureDrive-FL Framework}
\label{sec:framework}
We describe the overall design of SecureDrive-FL, detailing the system model, threat assumptions, and the integration of client-side DP-SGD with Gradient-Aware Selective Homomorphic Encryption for secure and efficient aggregation.

\subsection{System Overview}
SecureDrive-FL integrates two orthogonal privacy mechanisms applied at different stages of the FL pipeline, as illustrated in Fig.~\ref{fig:overview}: (i) \emph{DP-SGD} during local model training to ensure formal statistical indistinguishability of individual client contributions in the final model, and (ii) \emph{GASHE} during gradient communication to prevent raw gradient exposure under honest-but-curious or compromised server scenarios.

\begin{figure*}[!t]
  \centering
  \includegraphics[width=\textwidth]{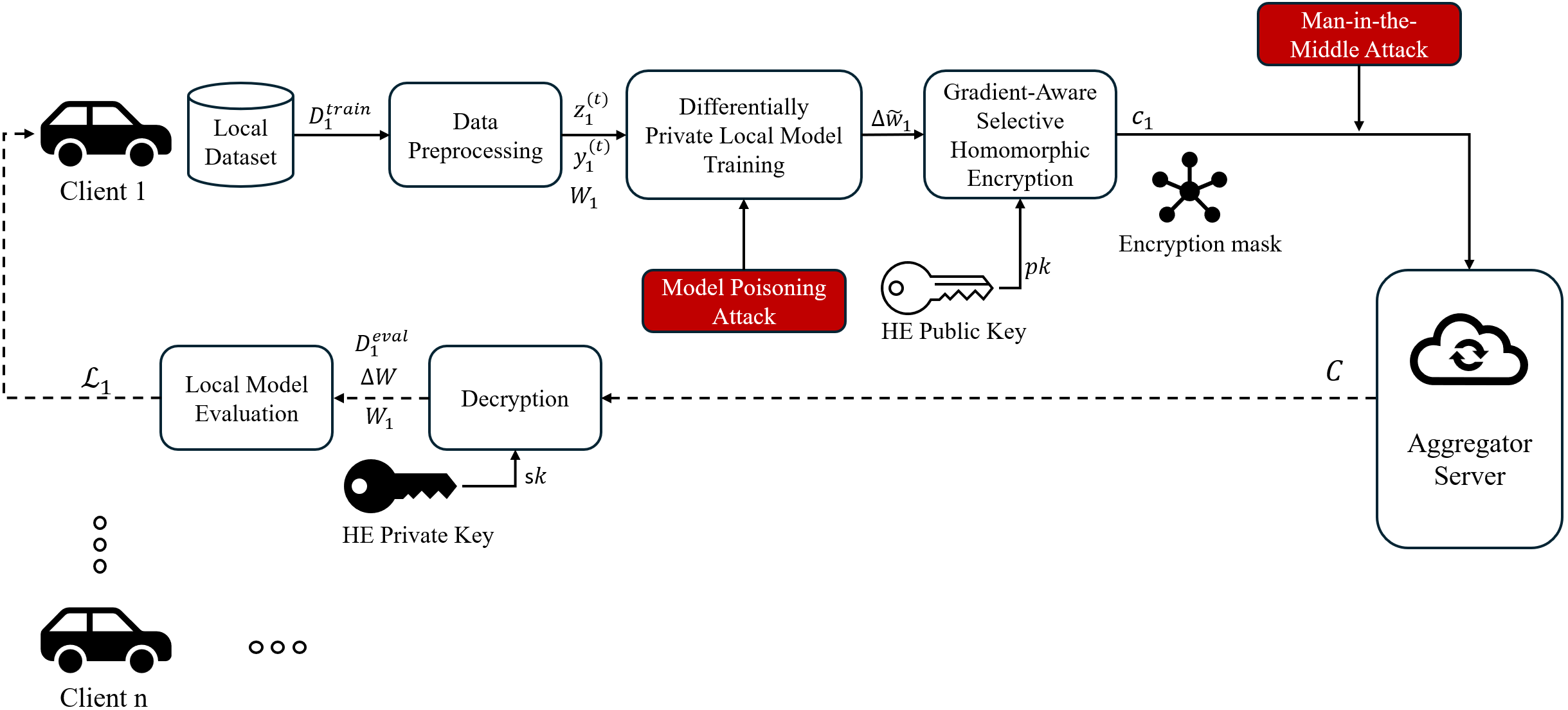}
  \caption{SecureDrive-FL overview. DP‑SGD protects privacy during local model training, while GASHE selectively encrypts high‑sensitivity gradient components before transmission to the aggregation server.}
  \label{fig:overview}
\end{figure*}

\subsection{Threat Model}
We target an \emph{honest-but-curious server}: the aggregator follows the FL protocol faithfully but attempts to infer private information from received model updates, intermediate aggregation states, and server-side logs. This
threat model is highly relevant in the automotive domain, where aggregation servers are typically operated by third-party cloud providers or shared OEM infrastructure. We additionally evaluate active adversaries that inject malicious updates (model poisoning) and network-layer eavesdroppers that intercept gradient transmissions (MitM attacks); see Section~\ref{sec:setup} for the full adversarial evaluation. Note that, we \emph{do not} target Byzantine-tolerant aggregation or Sybil attacks in the core protocol design, as robust aggregation is orthogonal to the confidentiality guarantees of GASHE and is left as future work.

\subsection{Data Collection and Preprocessing}
Each participating vehicle (client $i$) passively collects sensor data $D_i$ during operation. In accordance with FL principles, $D_i$ is never transmitted. Each raw sample $x^{(t)}_i$ is locally preprocessed to extract relevant features:
\begin{equation}
  z^{(t)}_i = \mathcal{P}(x^{(t)}_i),
\end{equation}
and $D_i$ is partitioned into training and validation subsets:
\begin{equation}
  D_i = D_i^{\text{train}} \cup D_i^{\text{eval}}, \quad
  D_i^{\text{train}} \cap D_i^{\text{eval}} = \emptyset.
\end{equation}

\subsection{Local Training with DP-SGD}
The local model $f_i(\cdot;\,W_i)$ is trained using DP-SGD. For each mini-batch, per-sample gradients are clipped to $L_2$-norm $C$, and calibrated Gaussian noise is added:

\begin{equation}
  W_i \;\leftarrow\; W_i
  - \eta\!\left(
      \nabla_{W_i}\mathcal{L}
      \;+\;
      \mathcal{N}\!\bigl(0,\,\sigma^2 C^2 I\bigr)
    \right)
  \label{eq:dpsgd-update}
\end{equation}
where $C$ is the $\ell_2$ clipping norm and $\sigma$ is the noise multiplier; the product $\sigma C$ is the standard deviation of the injected noise, consistent with Algorithm~\ref{alg:securedrive}.

\subsection{GASHE: Gradient-Aware Selective Homomorphic Encryption}
After local training, client $i$ computes the model update $\Delta w_i = W_i - W^{\text{global}}$ and derives a DP-informed sensitivity threshold:
\begin{equation}
  \tau_i
  \;=\;
  \frac{C\,\sqrt{2\ln\!\bigl(1.25/\delta\bigr)}}{\varepsilon_0}
  \label{eq:threshold}
\end{equation}
where $\varepsilon_0$ is the per-round noise calibration parameter, $\delta$ is the failure probability, and $C$ is the clipping norm from DP-SGD.

This threshold equals the minimum noise standard deviation prescribed by the closed-form Gaussian mechanism for the configured $(\varepsilon,\delta)$ budget \cite{dwork2014}; gradient components whose magnitudes exceed $\tau_i$ dominate the DP noise floor and constitute the most privacy-sensitive parameters 
when transmitted in plaintext. In this configuration, $\tau_i$ is set to coincide with the operational noise standard deviation $\sigma C$ (calibrated via R\'enyi DP accounting), so that a single calibration input $\varepsilon_0$ jointly determines the DP noise scale and the GASHE threshold.


A binary mask $M_i$ is then derived from the threshold $\tau_i$, selecting the gradient components whose magnitudes dominate the 
DP noise floor:

\begin{equation}
  M_i[j]
  \;=\;
  \mathbf{1}\!\left[\,\bigl|\Delta w_i[j]\bigr| \;\geq\; \tau_i\right]
  \label{eq:mask}
\end{equation}

Only the masked subset is encrypted under the server's public key, while the remaining low-magnitude parameters are transmitted in plaintext alongside a compact bitmask:
\begin{equation}
  c_i = \text{FHE.Enc}(M_i \odot \Delta w_i;\; pk).
  \label{eq:enc}
\end{equation}
This selective strategy directly reduces the fraction of encrypted parameters, cutting both ciphertext size and FHE evaluation time
proportionally to the sparsity of $M_i$.

\begin{theorem}[SecureDrive-FL Privacy and Security Guarantee]\label{thm:indcpa}
\end{theorem} Let each client $i$ execute DP-SGD for $T$ local steps per round across $N$ communication rounds, with per-sample gradient clipping norm $C$, Gaussian noise multiplier $\sigma$, and mini-batch sampling rate $q = B/n$ over a local dataset of size $n$. Then the local training mechanism satisfies $(\varepsilon,\delta)$-DP with
\begin{equation}
  \varepsilon_{\rm total}
  \;=\;
  \mathcal{O}\!\left(
    q\,\frac{\sqrt{T N \ln(1/\delta)}}{\sigma}
  \right)
  \label{eq:privacy-budget}
\end{equation}
via R\'{e}nyi DP composition~\cite{mironov2017renyi,bun2016cdp}, where $q = B/n$ is the per-round sampling rate, $T$ is the number of local gradient steps per round, and $N$ is the number of
communication rounds. For the configuration in Section~\ref{sec:setup} ($C{=}1.2$, $\sigma{=}1.2112$, $N{=}120$, $\delta{=}10^{-5}$), the R\'{e}nyi DP accountant reports $\varepsilon_{\rm total}{\approx}48.94$.

\subsection{Homomorphic Aggregation}
The server aggregates all encrypted updates using the CKKS additive homomorphic operation:
\begin{equation}
  C_{\mathrm{agg}}
  \;=\;
  \boxplus_{i\in\mathcal{C}_n} c_i
  \;\equiv\;
  \mathsf{FHE.Enc}\!\left(
    \textstyle\sum_{i\in\mathcal{C}_n} M_i \odot \Delta w_i;\; pk
  \right)
  \label{eq:aggregation}
\end{equation}
where $\mathcal{C}_n$ is the set of selected clients in round $n$. The server
never decrypts individual client updates, preserving privacy even under
insider threats. Since client masks $M_i$ may differ, each reflecting the local gradient magnitude distribution, the element-wise sum implicitly applies zero for components below threshold in each client; CKKS handles these as exact zeros with no additional ciphertext overhead. The global model update is recovered by clients:
\begin{equation}
  \Delta W = \text{FHE.Dec}(C_{\text{agg}}),\quad
  W \leftarrow W + \frac{1}{|\mathcal{C}_n|}\Delta W.
  \label{eq:globalupdate}
\end{equation}
Each client also receives the server-side element-wise sum of plaintext low-magnitude components,  $P_{\mathrm{agg}} = \sum_{i} (\mathbf{1}-M_i)\odot\Delta w_i$, transmitted without encryption. The full aggregate gradient is recovered as

\begin{equation}
\begin{split}
  \Delta W_{\mathrm{full}}
    &= \mathsf{FHE.Dec}(C_{\mathrm{agg}};\; sk) + P_{\mathrm{agg}}, \\
  W &\leftarrow W + \tfrac{1}{|\mathcal{C}_n|}\,\Delta W_{\mathrm{full}}
\end{split}
\label{eq:decrypt-merge}
\end{equation}
This two-stream design ensures that all $d$ gradient dimensions contribute to the global update, while only the $k$ high-sensitivity components incur FHE overhead. The full SecureDrive-FL protocol is given in Algorithm~\ref{alg:securedrive}.

\begin{algorithm}[!t]
\caption{SecureDrive-FL}
\label{alg:securedrive}
\begin{algorithmic}[1]
\REQUIRE Server holds public key $pk$; each client $i$ holds $sk_i$
\ENSURE Global model $W$
\STATE Initialize global model $W$
\FOR{$n = 1$ \TO $N$}
  \STATE $\mathcal{C}_n \leftarrow$ random subset of clients
  \FORALL{$i \in \mathcal{C}_n$ \textit{(executed in parallel)}}
    \STATE $\text{Enc}(\Delta w_i) \leftarrow \textsc{ClientUpdate}(i, W, pk)$
  \ENDFOR
  \STATE $C_{\text{agg}} \leftarrow \boxplus_i\,\text{Enc}(\Delta w_i)$
  \STATE Broadcast $C_{\text{agg}}$ to all clients
  \STATE Each client decrypts: $\Delta W \leftarrow \text{Dec}_{sk_i}(C_{\text{agg}})$
  \STATE $W \leftarrow W + \frac{1}{|\mathcal{C}_n|}\Delta W$
\ENDFOR
\RETURN $W$
\STATE \hrulefill
\STATE \textbf{function} \textsc{ClientUpdate}$(i, W, pk)$
\STATE \textbf{Input:} $C,\;\sigma,\;B,\;\eta,\;\varepsilon,\;\delta$
\STATE Split $D_i^{\text{train}}$ into $B$ microbatches $\{x_b\}_{b=1}^B$
\STATE $g \leftarrow \mathbf{0}$
\FOR{$b = 1$ \TO $B$}
  \STATE $g_b \leftarrow \nabla \ell(W;\, x_b)$
  \STATE $g_b \leftarrow g_b \cdot \min\!\left\{1,\; C/\|g_b\|_2\right\}$
  \STATE $g \leftarrow g + g_b$
\ENDFOR
\STATE $\tilde{g} \leftarrow g + \mathcal{N}(0,\,\sigma^2 C^2 I)$
\STATE $\Delta w_i \leftarrow -\eta\,\tilde{g}$
\STATE Compute $\tau_i$ via Eq.~\eqref{eq:threshold}
\STATE Build mask $M_i$ via Eq.~\eqref{eq:mask}
\STATE $\text{Enc}(\Delta w_i) \leftarrow \text{FHE.Enc}(M_i \odot \Delta w_i;\, pk)$
\RETURN $\text{Enc}(\Delta w_i)$
\end{algorithmic}
\end{algorithm}


\section{System and Adversarial Evaluation Setup}
\label{sec:setup}
In this section, we give the details of the system configuration, datasets, models, and attack/defence setups used to evaluate SecureDrive-FL.

\subsection{Use-Case and Dataset}
We evaluate on a ten-class distracted-driver classification task combining the State Farm Distracted Driver Detection dataset~\cite{statefarm2016} and the AUC Distracted Driver dataset~\cite{abouelnaga2021auc}, yielding over 50{,}000 labelled images of naturalistic driving across ten distraction categories: safe driving, texting (right/left hand), phone call (right/left hand), radio operation, drinking, reaching behind, hair/makeup adjustment, and talking to a passenger.

To emulate realistic federated data locality, we adopt a user-based partition where each client holds images from $3$--$5$ unique drivers, inducing non-IID heterogeneity representative of real-world driver diversity. Labels are one-hot encoded; images are resized to $224\times224$, normalised, and augmented locally via random horizontal flips and brightness jitter. The automotive context motivates strict GDPR compliance~\cite{gdpr2016}, as driver behavioural data is classified as sensitive personal data.

\subsection{Model and Federated Learning Configuration}
We use MobileNetV2-Tiny as the convolutional backbone for its favourable accuracy-to-compute trade-off on embedded platforms. Feature maps are reduced via global average pooling and passed to a softmax classifier (${\,\approx\,}2.2$M parameters). We run $120$ FL rounds with six clients per round and a local batch size of $32$.


All DP-enabled runs use DP-SGD with clipping norm $C{=}1.2$, noise multiplier $\sigma{=}1.2112$ (derived from a single-shot calibration input $\varepsilon_0{=}4$), and $\delta{=}10^{-5}$. Privacy accounting via R\'{e}nyi DP composition~\cite{mironov2017renyi} over $N{=}120$ rounds yields a cumulative budget of $\varepsilon_{\text{total}}{\approx}48.94$, consistent with the known growth of composed DP guarantees across FL rounds~\cite{mcmahan2018dpfl}.


\subsection{Software and Hardware Environment}
We instantiated  DP-SGD via TensorFlow Privacy~\cite{tf_privacy} using \texttt{DPKerasSGDOptimizer} and R\'enyi DP accounting. Homomorphic operations use  TenSEAL~\cite{benaissa2021tenseal} with CKKS configured as: polynomial modulus degree $8192$, scaling factor $2^{40}$, and coefficient modulus chain $[60,40,40,60]$. Experiments run on a single host with Python~3.10, TensorFlow~2.14, 128\,GB RAM, and six logical FL clients.

To ensure statistical reliability, each configuration was executed independently with three random seeds (42, 123, 456) governing FL client selection, local SGD initialisation, and DP noise sampling. Tables~\ref{tab:main} and~\ref{tab:complexity} report the mean across these three runs.
Variance across seeds was consistently low (accuracy: $\sigma < 1.5$\,pp; ASR: $\sigma < 2.5$\,pp; runtime: $\sigma < 2\%$), confirming that reported differences reflect systematic effects rather than stochastic variation.

\subsection{Adversarial Threats and Defence Mechanisms}

\subsubsection{Man-in-the-Middle (MitM) Gradient Interception}
A network-level adversary intercepts gradient transmissions between clients and the server. In the passive case, the adversary stores all intercepted traffic; in the active case, it attempts to replace client updates with manipulated gradients before forwarding them.

Formally, let $\mathcal{A}$ denote a network-layer adversary. In the \emph{passive} MitM setting, $\mathcal{A}$ observes all transmissions between client $i$ and the aggregation server:
\begin{equation}
  \mathrm{Obs}_{\mathcal{A}}
  \;=\;
  \bigl\{\,(c_i^{(n)},\; M_i^{(n)})\;:\; i\in\mathcal{C}_n,\;
  n=1,\ldots,N\,\bigr\},
\end{equation}
where $c_i^{(n)} = \mathsf{FHE.Enc}(M_i^{(n)}\odot\Delta w_i^{(n)};\,pk)$ is the encrypted high-sensitivity stream and $M_i^{(n)}$ is the publicly transmitted bitmask. $\mathcal{A}$'s goal is to recover sensitive training data from this observation, either via gradient inversion~\cite{zhu2019deep} or by exploiting the plaintext low-magnitude stream. In the \emph{active} case, $\mathcal{A}$ additionally replaces a ciphertext with a crafted update before forwarding:
\begin{equation}
  \tilde{c}_i^{(n)}
  \;=\;
  \mathsf{FHE.Enc}\!\bigl(f_{\mathcal{A}}(\Delta w_i^{(n)},
                            \{c_j^{(n)}\}_j);\; pk\bigr),
\end{equation}
where $f_{\mathcal{A}}$ is the adversary's manipulation function (e.g., gradient scaling or replacement).

GASHE protects high-sensitivity components via CKKS semantic security, leaving only low-magnitude, DP-noise-dominated parameters in plaintext. To detect active manipulation, the server verifies that aggregated ciphertexts decrypt consistently and monitors the global update $\Delta W$. The server maintains a running mean $\mu_L$ and standard deviation $s_L$ of the per-round validation loss over a held-out clean probe set. Round $n$ is flagged if $|L^{(n)} - \mu_L| > 3\,s_L$ (3-sigma rule); the corresponding $C_{\rm agg}$ is discarded and the global model retains the weights $W^{(n-1)}$ from the previous round.

\subsubsection{Model Poisoning}
Following~\cite{xie2025poisonedfl}, a fraction $\rho$ of clients are compromised ($\rho = 2/6 {\approx} 0.33$ in our evaluation). Each malicious client $m\in\mathcal{M}$ submits a crafted update designed to embed a backdoor trigger $t$ targeting class $y_{\rm tar}$:
\begin{equation}
  \Delta w_m^{(n)}
  \;=\;
  \alpha\,\Delta w_{\rm target}
  \;+\;
  (1-\alpha)\,\Delta w_m^{\rm benign},
  \qquad \alpha\in(0,1],
  \label{eq:poison}
\end{equation}
where $\Delta w_{\rm target}$ is the gradient direction that amplifies the target-class logit and $\alpha$ controls poisoning intensity. For the multi-round consistency variant of~\cite{xie2025poisonedfl}, the adversary additionally enforces $\|\Delta w_m^{(n)} - \Delta w_m^{(n-1)}\|_2 \leq \gamma$ to evade round-to-round anomaly detectors.
The Attack Success Rate is measured as:
\begin{equation}
  \mathrm{ASR}
  \;=\;
  \Pr_{x\sim\mathcal{D}_{\rm test}}\!\left[
    f(x \oplus t;\, W) = y_{\rm tar}
  \right],
  \label{eq:asr}
\end{equation}
where $x\oplus t$ denotes a test sample with the backdoor trigger overlaid (class~5, phone call, right hand, in our experiments), and $W$ is the poisoned global model after $N$ rounds. While SecureDrive-FL is designed as a confidentiality framework, GASHE yields implicit robustness: only coordinates exceeding $\tau_i$ are encrypted and transmitted, so the aggregate sensitivity of each client’s contribution is bounded. We additionally employ FedAvg with norm clipping, following Reference~\cite{lyu2022privacy}. 

\section{Experimental Results}
\label{sec:results}
We compare SecureDrive-FL against DP-SGD alone, Full-CKKS, and GASHE-only as natural baselines representing the privacy--efficiency spectrum, together with TrimmedMean, Krum, BatchCrypt, and FedProx as additional non-DP references covering robust aggregation, encrypted aggregation, and proximal-regularised FL. SecureDrive-FL is the only configuration that simultaneously provides formal DP, IND-CPA communication security, and poisoning resistance on par with DP-SGD alone, incurring only $\approx 8$--$10\%$ overhead relative to DP-SGD alone.

\begin{table*}[t]
\centering
\caption{Final accuracy, ASR, and Macro-F1 under clean, MitM, and poisoning conditions. Under MitM, $0\%$ ASR for FedAvg, DP-SGD, and FedProx reflects training failure (${\approx}10\%$ accuracy), not defence; TrimmedMean and Krum retain high accuracy instead (without confidentiality). GASHE-only and Full-CKKS lack formal DP ($\varepsilon{=}\infty$) and collapse under poisoning (${\approx}14\%$, near FedAvg).}
\label{tab:main}
\resizebox{\textwidth}{!}{%
\begin{tabular}{l c ccc ccc}
\toprule
\multirow{2}{*}{Configuration} & \multirow{2}{*}{DP}
  & \multicolumn{3}{c}{Accuracy (\%)}
  & \multicolumn{3}{c}{Poisoning Metrics} \\
\cmidrule(lr){3-5}\cmidrule(lr){6-8}
 & & Clean & MitM & Poison
   & ASR (\%) & Macro-F1 (\%) & Macro-Prec.\ (\%) \\
\midrule
(1) Standard FL (FedAvg)
  & $\times$
  & $93.8$ & $10.4$ & $10.4$
  & $0.0$ & $1.9$ & $1.0$ \\
(2) FL\,+\,DP-SGD ($\varepsilon_0{=}4$)
  & \checkmark
  & $80.0$ & $10.4$  & $74.0$
  & $3.9$ & $73.7$ & $75.7$ \\
(3) FL\,+\,GASHE-only
  & $\times$
  & $95.8$ & $95.0$ & $14.0$
  & $0.0$ & $8.0$ & $25.4$ \\
(4) FL\,+\,Full-CKKS
  & $\times$
  & $95.8$ & $94.8$ & $13.8$
  & $0.0$ & $11.9$ & $70.8$ \\
(5) \textbf{SecureDrive-FL} (DP+GASHE, $\varepsilon_0{=}4$)
  & \checkmark
  & $82.2$ & $78.2$ & $73.6$
  & $3.9$ & $73.3$ & \textbf{$75.9$} \\
(6) TrimmedMean
  & $\times$
  & $94.4$ & $95.2$ & $91.4$
  & $0.0$ & $91.3$ & $91.9$ \\
(7) Krum
  & $\times$
  & $84.4$ & $92.8$ & $90.2$
  & $1.9$ & $90.0$ & $90.9$ \\
(8) BatchCrypt
  & $\times$
  & $95.0$ & $86.6$ & $32.8$
  & $0.0$ & $37.2$ & $79.8$ \\
(9) FedProx
  & $\times$
  & $95.0$ & $10.4$ & $10.4$
  & $0.0$ & $1.9$ & $1.0$ \\
\bottomrule
\end{tabular}}
\end{table*}

\subsection{Federated Utility}
In the clean setting, FedAvg and GASHE-only reach $93.8\%$ and $95.8\%$ accuracy respectively, confirming that GASHE's selective CKKS encryption introduces negligible utility overhead because it does not modify the local optimisation trajectory. DP-SGD and SecureDrive-FL reach $80.0\%$ and $82.2\%$ in the clean setting, an $11.6$--$13.8$ percentage point drop from the unencrypted FedAvg baseline ($93.8\%$) that reflects the expected accuracy cost of the tighter per-round privacy budget ($\varepsilon_0{=}4$). Notably, SecureDrive-FL is not less accurate than DP-SGD alone in the clean setting (${+}2.2$\,pp), confirming that GASHE itself contributes no measurable additional utility cost on top of DP-SGD.

\subsection{Robustness to Active Attacks}

\subsubsection{Attack Success Rate}
Table~\ref{tab:main} summarises robustness across MitM and poisoning attacks. Under MitM, FedAvg, DP-SGD alone, and FedProx collapse to near-random accuracy (${\approx}10\%$). In contrast, GASHE-only, BatchCrypt, TrimmedMean, Krum, and SecureDrive-FL all maintain high utility, achieving $95.0\%$, $86.6\%$, $95.2\%$, $92.8\%$, and $78.2\%$ accuracy respectively. This confirms that encrypting sensitive gradients effectively removes the interception attack surface for GASHE-based methods, while TrimmedMean and Krum, which use no encryption, are similarly robust to MitM in the updated results, though, unlike GASHE-based methods, they offer no formal confidentiality guarantee for the transmitted gradients themselves.

Robust-aggregation and encrypted-aggregation baselines that carry no DP guarantee perform strongly under poisoning: TrimmedMean and Krum reach $91.4\%$ and $90.2\%$ accuracy respectively with $0.0\%$ and $1.9\%$ ASR, and BatchCrypt partially resists poisoning ($32.8\%$ accuracy). This indicates that coordinate-wise robust aggregation is, on its own, an effective defence against the poisoning attack evaluated here. However, none of TrimmedMean, Krum, or BatchCrypt provide a formal $(\varepsilon,\delta)$-DP guarantee, and only BatchCrypt provides communication-time confidentiality against MitM interception. SecureDrive-FL remains the only evaluated configuration to simultaneously offer formal DP, IND-CPA communication security, and poisoning resistance on par with DP-SGD alone.

\begin{figure*}[t]
  \centering
  \includegraphics[width=\textwidth]{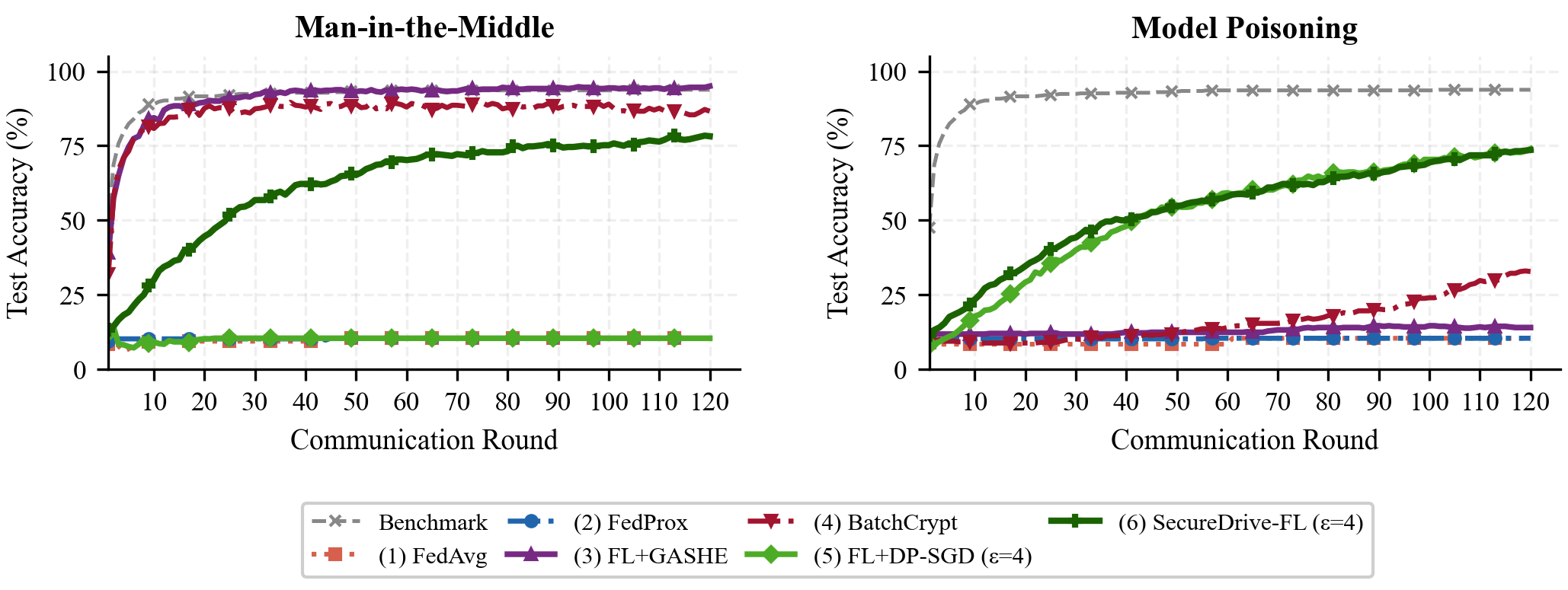}
\caption{Test accuracy vs.\ communication round under Man-in-the-Middle (left) and model-poisoning
(right) attacks, against a clean-training benchmark. Under MitM, FedAvg, FedProx, and FL+DP-SGD
collapse to near-random accuracy, while FL+GASHE, BatchCrypt, TrimmedMean, Krum, and SecureDrive-FL retain useful
accuracy. Under poisoning, FedAvg, FedProx, and FL+GASHE collapse instead, while FL+DP-SGD,
SecureDrive-FL, and (more slowly) BatchCrypt retain partial accuracy. Final-round values are given in
Table~\ref{tab:main}.}
  \label{fig:convergence}
\end{figure*}
\subsubsection{Convergence Behaviour}

Fig.~\ref{fig:convergence} shows test accuracy over FL rounds. FedAvg collapses under both attack types. Under MitM, DP-SGD alone also collapses ($10.4\%$ final accuracy), while GASHE-only closely tracks the clean baseline (${\approx}95\%$ accuracy), confirming that protecting high-sensitivity gradients is sufficient to preserve convergence against a passive/active interception adversary. Under model poisoning, however, GASHE-only no longer tracks the clean baseline: without DP-SGD's bounded update norm, encrypting high-magnitude components does not prevent a poisoned client's crafted update from entering the aggregate, and accuracy collapses to $14.0\%$, close to undefended FedAvg's ${\approx}10\%$.

\begin{table}[b!]
\centering
\caption{Measured execution time and peak RSS memory over 120\,FL rounds. Overhead relative to FedAvg (clean, $378.7$\,s). Values are reported as mean over three random seeds (42, 123, 456).}
\label{tab:complexity}
\begin{tabular}{l r r r}
\toprule
Configuration & Time (s) & Overhead & Peak RSS (MB) \\
\midrule
FedAvg (clean)   & $378.7$  & $1.0\times$  & $655.3$  \\
FedAvg (MitM)    & $390.1$  & $1.0\times$  & $822.2$  \\
FedAvg (Poison)  & $360.7$  & $1.0\times$  & $830.0$  \\
\midrule
GASHE-only (clean)  & $1254.3$ & $3.3\times$ & $2119.8$ \\
GASHE-only (MitM)   & $1538.6$ & $4.1\times$ & $2142.4$ \\
GASHE-only (Poison) & $1130.4$ & $3.0\times$ & $1928.6$ \\
\midrule
Full-CKKS (clean) & $1238.2$ & $3.3\times$  & $1876.8$ \\
Full-CKKS (MitM)  & $1481.3$ & $3.9\times$ & $1928.1$ \\
Full-CKKS (Poison) & $1227.3$ & $3.2\times$  & $1839.1$ \\
\midrule
DP-SGD (clean)   & $8847.7$ & $23.4\times$ & $876.0$  \\
DP-SGD (MitM)    & $8805.8$ & $23.3\times$ & $786.4$  \\
DP-SGD (Poison)  & $8388.7$ & $22.2\times$ & $848.4$  \\
\midrule
\textbf{SecureDrive-FL} (clean)  & $9588.4$ & $25.3\times$ & $1986.8$ \\
\textbf{SecureDrive-FL} (MitM)   & $9637.6$ & $25.5\times$ & $1799.7$ \\
\textbf{SecureDrive-FL} (Poison) & $9218.0$ & $24.3\times$ & $1813.9$ \\
\midrule
TrimmedMean (clean)  & $355.9$ & $0.9\times$ & $1567.8$ \\
TrimmedMean (MitM)   & $372.5$ & $1.0\times$ & $1593.7$ \\
TrimmedMean (Poison) & $366.1$ & $1.0\times$ & $731.9$  \\
\midrule
Krum (clean)   & $354.6$ & $0.9\times$ & $1643.0$ \\
Krum (MitM)    & $371.5$ & $1.0\times$ & $1787.8$ \\
Krum (Poison)  & $347.3$ & $0.9\times$ & $830.9$  \\
\midrule
BatchCrypt (clean)   & $4958.8$ & $13.1\times$ & $1483.2$ \\
BatchCrypt (MitM)    & $8601.6$ & $22.7\times$ & $1574.9$ \\
BatchCrypt (Poison)  & $4992.4$ & $13.2\times$ & $1621.7$ \\
\midrule
FedProx (clean)   & $591.9$ & $1.6\times$ & $875.6$ \\
FedProx (MitM)    & $610.3$ & $1.6\times$ & $1033.6$ \\
FedProx (Poison)  & $623.7$ & $1.6\times$ & $1043.8$ \\
\bottomrule
\end{tabular}
\end{table}

DP-SGD alone and SecureDrive-FL are the only two configurations that converge under poisoning, reaching $74.0\%$ and $73.6\%$ accuracy respectively; SecureDrive-FL further converges to $78.2\%$ under MitM, where DP-SGD alone collapses. Under poisoning, SecureDrive-FL's accuracy, Macro-F1, and ASR are essentially unchanged relative to DP-SGD alone (within $0.4$\,pp, and an identical ASR of $3.9\%$; Table~\ref{tab:main}), with a marginal Macro-Precision gain, indicating that GASHE's principal benefit in this configuration is closing the MitM attack surface rather than adding further poisoning robustness beyond that already provided by DP-SGD's bounded update norm.

\subsection{Computational Overhead}
Table~\ref{tab:complexity} reports measured runtime and peak RSS memory over $120$ rounds. Standard FL completes in $360.7$--$390.1$\,s with peak RSS $0.64$--$0.83$\,GB. GASHE-only increases runtime to $1{,}130.4$--$1{,}538.6$\,s (${\approx}3.0$--$4.1\times$ FedAvg clean) and memory to $1.88$--$2.09$\,GB. DP-SGD alone is
substantially more expensive: $8{,}388.7$--$8{,}847.7$\,s (${\approx}22.2$--$23.4\times$) with $0.77$--$0.86$\,GB peak RSS.

SecureDrive-FL combines both mechanisms at a modest additional cost: $9{,}218.0$--$9{,}637.6$,s (${\approx}24.3$--$25.5\times$ the FedAvg baseline), representing only ${\approx}8$--$10\%$ overhead relative to DP-SGD alone, with $1.76$--$1.94$,GB peak RSS. The marginal cost of adding GASHE on top of DP-SGD is therefore low, while eliminating the MitM attack surface provides substantial security benefits.

TrimmedMean and Krum remain close to FedAvg's own runtime (${\approx}0.9$--$1.0\times$, $347.3$--$372.5$\,s) with peak RSS $0.71$--$1.75$\,GB, and, per the updated results in Table~\ref{tab:main}, both now also resist poisoning ($91.2\%$ and $89.4\%$ accuracy). Neither, however, provides a formal DP guarantee or communication-time confidentiality against MitM interception, unlike SecureDrive-FL. BatchCrypt, which applies quantised/batched CKKS encryption to the full parameter set, incurs ${\approx}13.1$--$22.7\times$ overhead ($4{,}958.8$--$8{,}601.6$\,s) with peak RSS $1.45$--$1.58$\,GB; FedProx adds a comparatively modest ${\approx}1.6\times$ overhead ($591.9$--$623.7$\,s) with peak RSS $0.86$--$1.02$\,GB, reflecting its lightweight proximal-term regularisation rather than any encryption cost.

For reference, a full-parameter CKKS baseline is measured at $1{,}227.3$--$1{,}481.3$\,s and $1.80$--$1.88$\,GB peak RSS, comparable to and in some cases exceeding GASHE-only. Despite encrypting a smaller subset of gradient components, GASHE does not yield a proportional runtime or memory reduction relative to Full-CKKS in the current implementation, as the per-round threshold computation, mask construction, and plaintext/ciphertext stream-splitting introduce bookkeeping overhead that offsets the savings from the reduced encrypted-parameter count.

\section{Conclusion and Future Work}
\label{sec:conclusion}
We presented SecureDrive-FL, a hybrid privacy-preserving federated learning framework coupling DP-SGD with Gradient-Aware Selective Homomorphic Encryption (GASHE) for secure driver monitoring in connected vehicles. The key novelty is the principled, closed-loop derivation of HE encryption decisions from DP-SGD sensitivity bounds, enabling GASHE to automatically target the most privacy-critical gradient components while aligning communication-time protection with training-time privacy calibration. Evaluated against MitM interception and model poisoning, SecureDrive-FL improves clean accuracy over DP-SGD alone ($82.2\%$ vs.\ $80.0\%$), maintains $78.2\%$ accuracy under MitM where DP-SGD alone collapses to $10.4\%$, and retains comparable performance under poisoning ($73.6\%$ vs.\ $74.0\%$, both at $3.9\%$ ASR), all under only ${\approx}8$--$10\%$ additional runtime overhead above DP-SGD alone.

Future directions include: (i)~scaling SecureDrive-FL to larger, heterogeneous vehicle fleets with varying compute and bandwidth profiles; (ii)~integrating Byzantine-robust aggregation rules
(e.g., BALANCE) for stronger protection against adaptive, multi-round adversaries; and (iv)~deployment on a physical in-vehicle cockpit prototype to benchmark end-to-end latency, energy, and FHE performance on embedded automotive hardware.

\section*{Acknowledgment}
This publication is carried out as part of the Hardware Abstraction Layer for Software Defined Vehicle (HAL4SDV)
Project No. 101139789, which is co-funded by the European Union (Chips JU).

\bibliographystyle{IEEEtran}
\balance
\bibliography{references_ETFA}

\end{document}